# STATUS OF NON-DESTRUCTIVE BUNCH LENGTH MEASUREMENT BASED ON COHERENT CHERENKOV RADIATION *

J. B. Zhang, S. L. Lu, T. M. Yu, H. X. Deng[#], SINAP/SSRF, Shanghai, 201800, China
D. Shkitov, M. Shevelev, G. Naumenko, A. Potylitsyn, TPU, Tomsk, 634050, Russia

*Abstract*

As a novel non-destructive bunch length diagnostic of the electron beam, an experimental observation of the coherent Cherenkov radiation generated from a dielectric caesium iodide crystal with large spectral dispersion was proposed for the 30MeV femtosecond linear accelerator at Shanghai Institute of Applied Physics (SINAP). In this paper, the theoretical design, the experimental setup, the terahertz optics, the first angular distribution observations of the coherent Cherenkov radiation, and the future plans are presented.

## INTRODUCTION

In order to verify the theoretical models of short pulse electron bunch generation, determine and understand the outcomes of the free electron laser experiments, in past decades, significant efforts were devoted to characterize the temporal structure of the electron bunch. Since the electron bunch length is much shorter than the temporal resolution of standard electron beam diagnostics devices, various instrumentations such as transverse deflecting radiofrequency cavity [1], electro-optical sampling [2, 3] and optical replica synthesizer techniques [4-6] have been demonstrated for electron bunch length measurements. However, these time-domain techniques are usually costly and complicated.

An alternative way to measure the bunch length is characterizing those coherent radiations generated at wavelengths comparable to, or longer than, the electron bunch length, when all electrons in the bunch irradiate more or less in phase. The intensity of coherent radiation is proportional to the square of the bunch population and the coherent radiation spectrum contains the information about the current distribution of the electron bunch, e.g., coherent transition radiations [7] and coherent diffraction radiations [8] generated when the electron bunch passes and moves in the vicinity of a medium target respectively. Generally, these coherent radiations were extracted and spectrally investigated by terahertz interferometer, such as Mickelson and Martin-Puplett type in which the spectrum resolution is highly related on the terahertz beam splitter and the motion accuracy of the movable mirror. Moreover, it is hard to figure out terahertz spectrum in a single-shot for conventional interferometers.

Cherenkov radiation is emitted when a charged particle passes through a dielectric medium at a speed greater than the phase velocity of light in that medium. The charged particles polarize the molecules of that medium, which then turn back rapidly to their ground state, thus emitting Cherenkov radiation in the process. More recently, it is found that Cherenkov radiation can be emitted not only when the electron beam pass through a medium target, but also when it moves in vicinity of a medium target [9], which holds a promising prospect in non-invasive beam diagnostics in modern accelerator [10]. A non-destructive bunch length measurement test bench on the basis of coherent Cherenkov radiation from a dielectric caesium iodide (CsI) target with large spectral dispersion was proposed and established recently [11] for femtosecond linear accelerator [12] at Shanghai Institute of Applied Physics (SINAP). In this paper, the status of the bunch length measurement experiment is presented.

## PRINCINPLE DESCRIPTION

As illustrated in Fig. 1, when the electron bunch moves near a dielectric medium with velocity that exceeds the speed of light in this transparent medium, Cherenkov radiation will be emitted and the radiation cone in this case is defined by the following condition:

$$n\beta\cos\varphi = 1, \quad (1)$$

where $\varphi$ is the radiation angle, $\beta$ is the electron velocity in the speed of light units, $n$ is the refracted index.

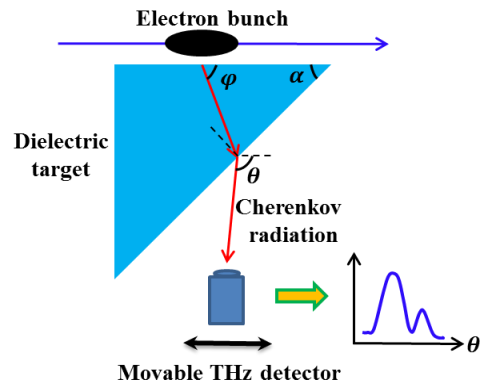

Figure 1: Scheme of coherent Cherenkov radiation based electron bunch measurement.

If a dielectric medium with large frequency dispersion was utilized, the different wavelength part of Cherenkov radiation would be radiated under different angles. Thus, one may change the complicated spectral measurements by more convenient angular ones in order to obtain bunch form-factor and the bunch length.

___________________
*Work supported by the joint Russian-Chinese grant (RFBR 110291177 and NSFC 11111120065) and partially by the Program of Russian MES "Nauka", National Natural Science Foundation of China (11175240), and Knowledge Innovation Program of Chinese Academy of Sciences.

#denghaixiao@sinap.ac.cn

To calculate the Cherenkov radiation observation angle dependence on wavelength, we have $d\theta/d\lambda = d\theta/dn \times dn/d\lambda$, where $\theta$ is the observation angle, $\lambda$ is the wavelength and $n$ is the refractive index. The refractive index $n$ of a particular transparent medium may be described by the Sellmeier formula, an empirical relationship of refractive index and radiation wavelength, reads

$$n(\lambda) = 1 + \sum \frac{A_i}{1 - B_i^2/\lambda^2[\mu m]} \quad (2)$$

A1=3.77, B1=9522, A2=1.33 and B2=26569 have been demonstrated for dielectric mediums of ceasium iodide in the terahertz region [13], which shows a large frequency dispersion and potential as a terahertz spectrometer. Then according to the Cherenkov criterion and the Snell's law, a maximum reasonable value of $d\theta/dn$ was modelled for target geometry with an optimal prism angle $\alpha$ about 45 degree for the target shown in Fig. 1. Considering the hygroscopy of a pure CsI, a Cherenkov target made of CsI with thallium admixture was manufactured and used for the bunch length experiment.

Using the universal model for describing polarization radiation [14], the theoretical angular distribution was calculated for the parameters of the SINAP femtosecond linear accelerator and the abovementioned characteristic of CsI Cherenkov target, as shown in Fig. 2. The SINAP femtosecond linear accelerator consists of a thermionic RF gun, an alpha magnet and a S-band accelerating tube, where the alpha magnet is used to compress the electron beam from few hundreds of ps to few hundreds of fs in longitudinal length and the SLAC-type accelerating tube is used to accelerate the electron beam up to 30MeV to minimize the lengthening force in the drift space. The design parameters of electron beam are listed in Table 1.

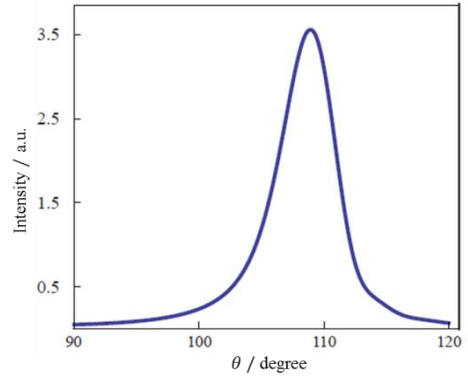

Figure 2: Theoretical angular distribution of the coherent Cherenkov radiation.

Table 1: Main parameters of electron beam

| Beam energy | 20 - 30 MeV |
|---|---|
| Beam charge | 0.068 nC |
| Normalized emittance | ~ 10 mm·mrad |
| Micro-bunch duration (FWHM) | 0.3 - 3 ps |

## EXPERIMENTAL SETUP

The experiment preparations include commissioning of the SINAP femtosecond linear accelerator, novel vacuum chamber equipped with various dielectric targets (i.e., CsI crystal prism for Cherenkov radiation and aluminum foil for transition/diffraction radiation) which can be remotely controlled by step motor system, And a terahertz beam line consists of a parabolic mirror (152.4mm focus length) and a terahertz detector is used to observe the angular distribution of coherent Cherenkov radiation.

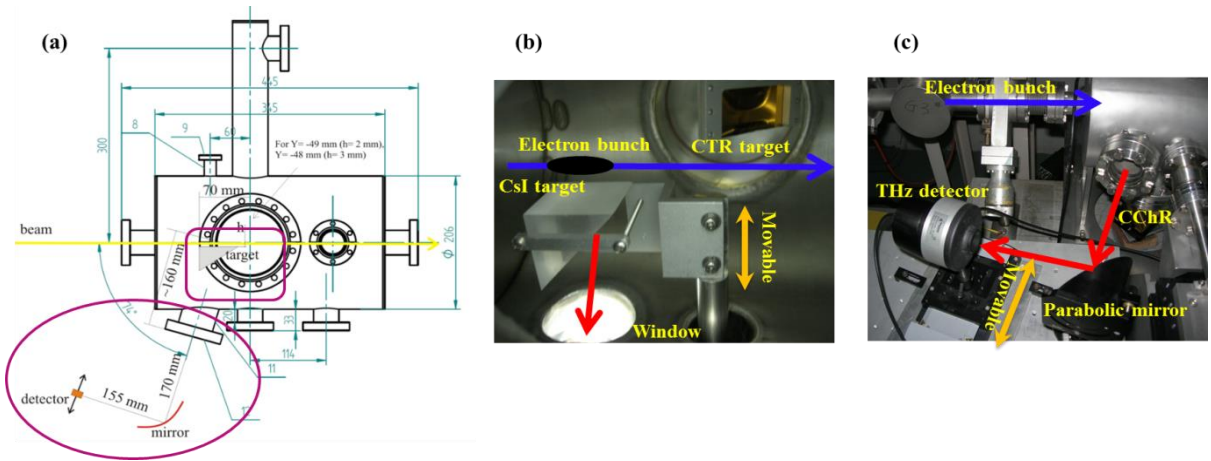

Figure 3: Experimental setup of the non-invasive bunch length measurement at SINAP femtosecond linear accelerator. (a) shows the schematic of all the devices involved, (b) shows an inner structure photo of the vacuum chamber where Cherenkov target and coherent transition radiation target coexist, (c) shows a photo of the terahertz optics for collecting the extracted coherent Cherenkov radiation. (b) and (c) represents the rectangular and ellipse part in (a), respectively.

As illustrated in Fig. 3, the CsI crystal target with sizes of 40mm×40mm×40mm was fixed on a rigid holder. The original position of the detector was supposed to be on an observation angle of 106 degree and the distance from the radiation source to the detector is 473mm.

## OBSERVED ANGULAR DISTRIBUTION

Before we start the experiment with the electron beam, the terahertz optics was firstly aligned by the He-Ne laser. To exclude the prewave zone effect [15], the detector was place and moved in the focal plane of the parabolic mirror. Then once the electron beam is ON, Cherenkov radiation signal is detected, as observed on oscilloscope in Fig. 4.

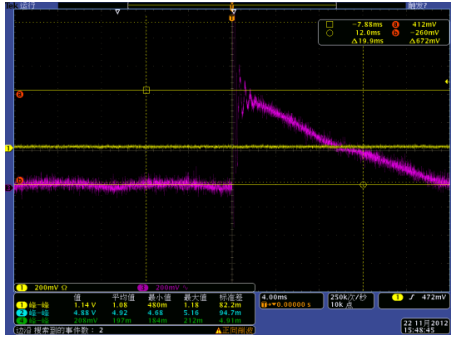

Figure 4: Coherent Cherenkov radiation signal observed on the oscilloscope.

Then we measure the angular distribution of coherent Cherenkov radiation by moving the terahertz detector perpendicularly to the incident Cherenkov light. In Fig. 5, the measured angular distribution is presented. In order to demonstrate the frequency dispersion of the experimental scheme, the current of alpha magnet was increased from 5.2A to 5.6A which will lengthen the electron bunch. The suppression of Cherenkov radiation yield was obtained, however, no obvious angular shift was observed. It may be attributed to the imperfectness of CsI target prism, high frequency suppression due to the transmission efficiency and limited aperture of the glass window. In further laser based THz-TDS measurement, the glass window results a huge absorption for terahertz radiation frequency higher than 0.3THz.

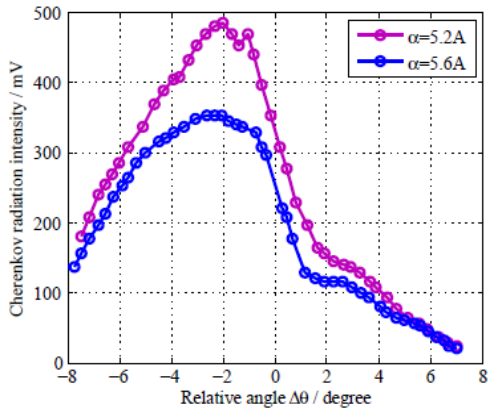

Figure 5: Experimental angular distribution of coherent Cherenkov radiation extracted from glass window.

A HDPE window with transmission efficiency larger than 50% for 0.2-10THz radiation was used to replace the glass window. Then we measure the angular distribution of Cherenkov radiation again. As shown in Fig. 6, when the current of alpha magnet was increased from 5.6A to 6.0A, the suppression of Cherenkov radiation intensity, together with an obvious angular shift of 0.8 degree were observed. It opens a possibility to estimate the bunch length by transferring the angular distribution of coherent Cherenkov radiation to the spectrum.

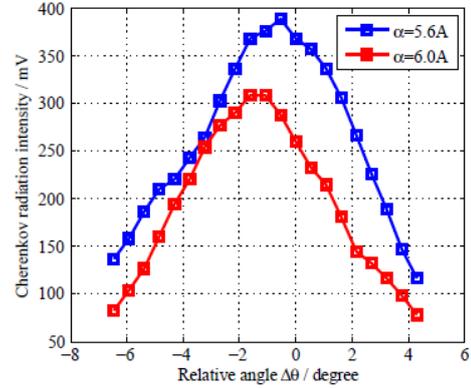

Figure 6: Experimental angular distribution of coherent Cherenkov radiation extracted from HDPE window.

## CONCULSIONS & FUTURE PLAN

As a non-destructive electron bunch length diagnostic, the angular distribution of coherent Cherenkov radiation generated from a dielectric caesium iodide crystal with large spectral dispersion was experimentally observed for the femtosecond linear accelerator at Shanghai Institute of Applied Physics. In this paper, the principle, experimental setup and the first experiment results are presented. In the future, calibrations between the spectral distribution and angular distribution of Cherenkov radiation is expected to be carried out, then with a detector array, the developed technique is capable for single-shot non-invasive bunch length measurement in the sub-picosecond range.

## ACKNOWLEDGMENT

The authors are grateful to D. K. Jiang, K. R. Ye, Y. B. Leng and W. Li from SSRF for their help in the vacuum chamber manufacture and related discussions.